\documentclass[useAMS,usenatbib]{mn2e}

\usepackage{graphicx}
\usepackage{amsmath}
\usepackage{amssymb}
\usepackage{color}
\usepackage{subfig}
\usepackage[breaklinks,colorlinks,citecolor=blue,linkcolor=red]{hyperref} 
\usepackage[all]{hypcap}

\newcommand\msun{\, \rm M_\odot}

\newcommand\gyr{{\, \rm Gyr}}

\newcommand\eout{{e_{\rm out}}}
\newcommand\eina{{e_{12}}}
\newcommand\einb{{e_{34}}}
\newcommand\aout{{a_{\rm out}}}

\newcommand\ain{a_{\rm in}}
\newcommand\aina{{a_{12}}}
\newcommand\ainb{{a_{34}}}
\newcommand\amax{{a_{\rm max}^{\rm out}}}

\title[BH mergers from quadruples]{Black hole mergers from quadruples}
\author[G. Fragione, B. Kocsis]{  \parbox{\textwidth}{Giacomo Fragione$^{1}$\thanks{E-mail: giacomo.fragione@mail.huji.ac.il} and Bence Kocsis$^{2}$}\\
{$^1$Racah Institute for Physics, The Hebrew University, Jerusalem 91904, Israel}\\
{$^2$Institute of Physics, E\"{o}tv\"{o}s University, P\'azm\'{a}ny P. s. 1/A, Budapest, 1117, Hungary}
}

\begin{document}

\maketitle

\begin{abstract}
With the hundreds of merging binary black hole (BH) signals expected to be detected by LIGO/Virgo, LISA and other instruments in the next few years, the modeling of astrophysical channels that lead to the formation of compact-object binaries has become of fundamental importance. In this paper, we carry out a systematic statistical study of quadruple BHs consisting of two binaries in orbit around their center of mass, by means of high-precision direct $N$-body simulations including Post-Newtonian (PN) terms up to 2.5PN order. We found that most merging systems have high initial inclinations and the distributions peak at $\sim 90^\circ$ as for triples, but with a more prominent broad distribution tail. We show that BHs merging through this channel have a significant eccentricity in the LIGO band, typically much larger than BHs merging in isolated binaries and in binaries ejected from star clusters, but comparable to that of merging binaries formed via the GW capture scenario in clusters, mergers in hierarchical triples, or BH binaries orbiting intermediate-mass black holes in star clusters. We show that the merger fraction can be up to $\sim 3$--$4\times$ higher for quadruples than for triples. Thus even if the number of quadruples is $20\%$--$25\%$ of the number of triples, the quadruple scenario can represent an important contribution to the events observed by LIGO/VIRGO.
\end{abstract}


\section{Introduction}

The LIGO-Virgo collaboration has recently released a catalogue of compact object mergers due to gravitational wave (GW) emission, comprised of ten merging black hole (BH) binaries and one merging neutron (NS) binary \citep{ligo2018}. As the detector sensitivity is improved, hundreds of merging binary signals are expected to be detected in the next few years, and the modeling of astrophysical channels that lead to the formation of compact-object binaries has become of fundamental importance.

Several astrophysical channels have been proposed leading to merging compact objects. Possibilities include isolated binary evolution through a common envelope phase \citep{bel16b,gimap18}, envelope expansion and fallback \citep{tag18} or chemically homogeneous evolution \citep{mand16,march16}, Lidov-Kozai (LK) mergers of binaries in galactic nuclei \citep{antoper12,petr17,hamer18,hoan18,fragrish2018,grish18} and in stellar triple systems \citep{ant17,sil17,alk2018,ll18}, mergers of wide binaries \citep{mich2019}, GW capture events in galactic nuclei \citep{olea09,rass2019}, mergers in active galactic nuclei (AGN) accretion disks \citep{bart17,sto17}, mergers in dark matter halos \citep{bird16,sas16}, and mergers in star clusters \citep{por00,olea06,askar17,baner18,cho18,frak18,rast2018,rod18}. While typically each model accounts for roughly the same rate ($\sim\ \mathrm{few}$ Gpc$^{-3}$ yr$^{-1}$, with the possible exception of the isolated binary case which predicts possibly much higher rates), the statistical contribution of different astrophysical channels can be hopefully disentangled using the spin, mass, eccentricity and redshift distributions \citep[see e.g.][]{olea16,gondan2018,samas18,zevin18}.

Bound stellar multiples are not rare. In particular, for massive stars which are progenitors of NSs and BHs,
observations have shown that more than $\sim 50$\% and $\sim 15$\% have at least one or two stellar companions, respectively \citep{duq91,ragh10,sa2013AA,tok14a,tok14b,duns2015,sana2017,jim2019}. Most previous dynamical studies on bound multiples have focused on determining the BH merger rate from isolated bound triples or triple BHs in globular clusters \citep{wen03,antcha2016,ant17,sil17,alk2018}. Quadruple systems are also observed. \citet{rid15} found a $\sim 4\%$ abundance  of $2$+$2$ quadruples. Compared to triple systems, quadruples have six additional degrees of freedom (i.e. the Keplerian elements) and excursions to very high eccentricity can take place over a much larger fraction of the parameter space \citep{pejcha2013}.

The dynamics of 2+2 quadruple systems has been under scrutiny to investigate a number of astrophysical phenomena, including orbital synchronization \citep{set18} and white dwarf-white dwarf mergers \citep{fan17,hamers2018}. \citet{vok2016} and \citet{breit2018} studied the secular and resonant dynamics of 2+2 systems, while \citet{alk2018} and \citet{zevin18} examined binary-binary scattering events, which may lead to eccentric BH mergers. Recently, \citet{liu2018} have discussed the possibility that the richer dynamics of quadruples can lead to a merger fraction $\sim 10$ times higher than that of triple systems, thus highlighting the relevance of this channel.

In this paper, we study for the first time the dynamical evolution of 2+2 quadruple BHs by means of direct high-precision $N$-body simulations, including Post-Newtonian (PN) terms up to 2.5PN order. In our calculations, we consider a power-law mass spectrum for the BHs, different maximum extensions of the quadruple and different distributions of the inner and outer semi-major axes and eccentricities. We quantify how the probability of merger depends on the initial conditions and determine the parameter distribution of merging systems relative to the initial distributions.

The paper is organized as follows. In Section~\ref{sect:kozai}, we discuss the relevant timescales of the considered systems. In Section~\ref{sect:mergers}, we present our numerical methods to determine the rate of BH mergers in quadruples, and discuss the parameters of merging systems. Finally, in Section~\ref{sect:conc}, we discuss the implications and draw conclusions.

\section{Lidov-Kozai mechanism in quadruples}
\label{sect:kozai}

\begin{figure} 
\centering
\includegraphics[scale=0.425]{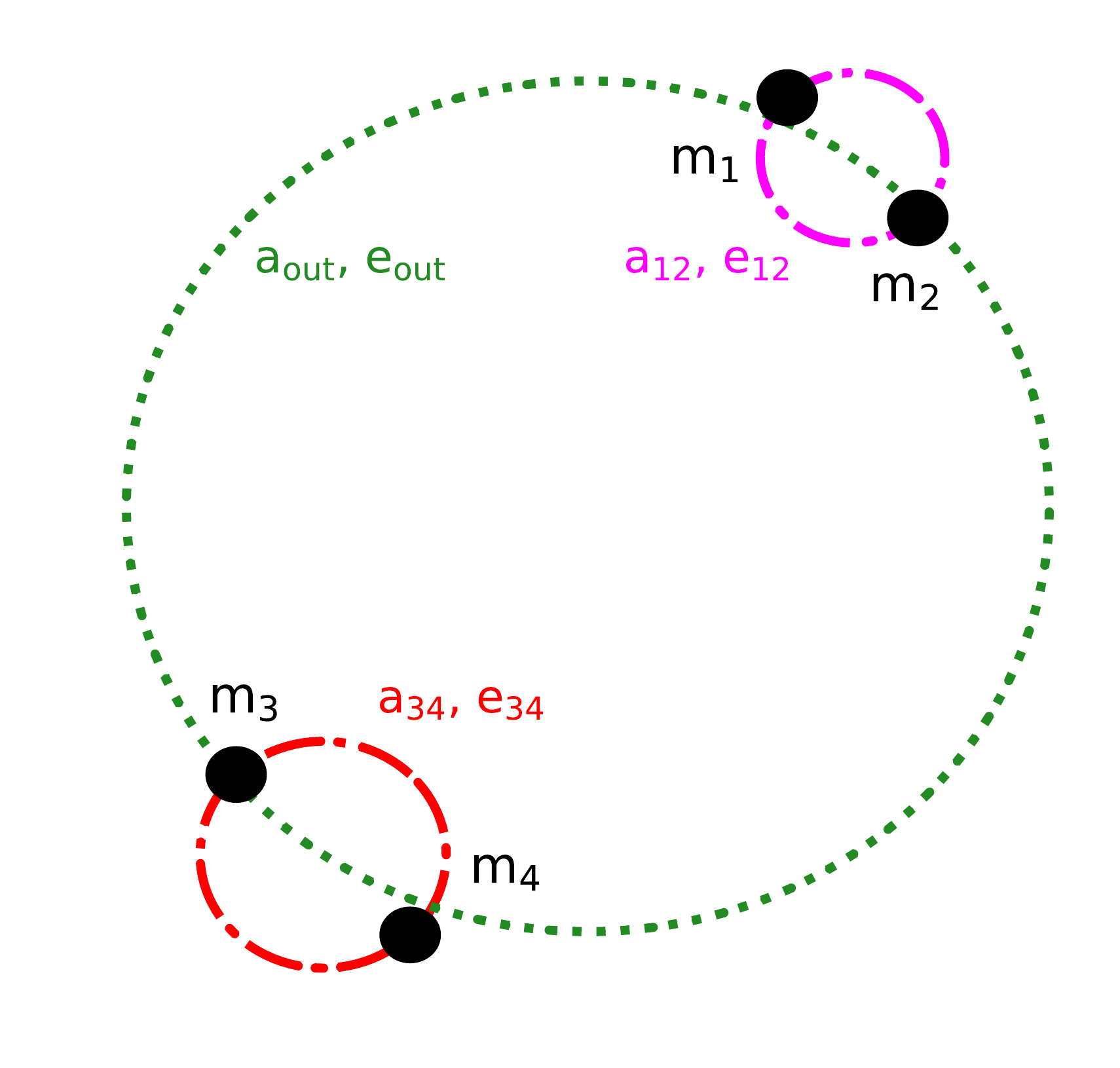}
\caption{The four-body system studied in the present work. We denote the mass of the components of the first inner binary as $m_{1}$ and $m_{1}$ and of the second inner binary as $m_{3}$ and $m_{4}$. The semimajor axis and eccentricity of the inner binaries are $\aina$, $\eina$ and $\ainb$, $\einb$, respectively, and of the outer orbit $\aout$ and $\eout$.}
\label{fig:fourbody}
\end{figure}

Figure~\ref{fig:fourbody} illustrates the system considered in this paper. The system consists of two inner binaries with masses $(m_{1}, m_{2})$ and $(m_{3}, m_{4})$, respectively, whose center of masses orbit around the 4-body system's center of mass, denoted as the outer binary. The semimajor axes and eccentricities of the inner binaries are denoted by $\aina$, $\eina$ and $\ainb$, $\einb$, respectively, and of the outer orbit by $\aout$ and $\eout$. 

In the case of a triple system made up of an inner binary that is orbited by a companion, the inner eccentricity and inclination oscillate due to the quadrupole moment of the tidal potential of the third body via the LK mechanism whenever the initial mutual orbital inclination is between $i_0\sim 40^\circ$--$140^\circ$ \citep{lid62,koz62}. The semimajor axis is approximately fixed during these oscillation cycles. At the secular quadrupole order of approximation, these oscillations occur on a timescale \citep{nao16}
\begin{equation}
T_{\rm LK}=\frac{8}{15\pi}\frac{m_{\rm tot}}{m_{\rm 3b}}\frac{P_{\rm 3b}^2}{P_{\rm in}}\left(1-e_{\rm 3b}^2\right)^{3/2}\ ,
\end{equation}
where $P_{{\rm in}}$ and $P_{{\rm 3b}}$ are the orbital periods of the inner and outer binary, respectively, $m_{\rm 3b}$ is the mass of the perturber, and $m_{\rm tot}$ is the total mass of the triple system. At the quadruple order of approximation, the maximal eccentricity of the inner binary due to the LK mechanism depends on the initial mutual inclination
\begin{equation}
e_{\rm in}^{\rm max}=\sqrt{1-\frac{5}{3}\cos^2 i_0}\ .
\label{eqn:emax}
\end{equation}

As $i_0$ approaches $\sim 90^\circ$, the inner binary eccentricity approaches almost unity. In the case of a compact-object binary, the large eccentricity makes its GW merger time shorter since it dissipates energy more efficiently at the pericentre \citep[e.g., see][]{antognini15,nao16}. Nevertheless, LK cycles can be suppressed by additional sources of precession, such as tidal bulges or general relativistic precession, which operate on a typical timescale of \citep{nao16}
\begin{equation}
T_{\rm GR}=\frac{a_{\rm in}^{5/2}c^2(1-e_{\rm in}^2)}{3G^{3/2}(m_{\rm in,1}+m_{\rm in,2})^{3/2}}\ .
\end{equation}
If $T_{\rm GR}<T_{\rm LK}$, the LK oscillations of the $(e_{\rm in},i)$ orbital elements are damped by relativistic effects \citep{naoz13}.

If the octupole corrections are taken into account, the eccentricity excitation becomes more prominent. In the case the outer orbit is eccentric, the inner eccentricity can reach almost unity even if the initial inclination is outside of the $i_0\sim 40^\circ$-$140^\circ$ Kozai-Lidov range \citep{naoz13a,li14}.

In the case the third companion is itself a binary star, as assumed in this paper, each binary acts as a distant perturber inducing LK cycles on the other binary. The evolution of such 2+2 quadruple systems differs from a combination of two uncoupled three-body LK processes. \citet{pejcha2013} showed that the binaries can experience coherent eccentricity oscillations and excursions to very high eccentricity that take place over a much larger fraction of the parameter space compared to triple systems. \citet{hamerslai2017} showed that the reason for this behaviour is that the second binary can induce nodal precession to the outer (2+2) orbit, which can cause a secular resonance if it happens on a timescale comparable to the LK timescale, and make the first binary merge. With respect to the first inner binary, the strength of this effect is parametrised by the dimensionless parameter \citep{hamerslai2017}
\begin{equation}
\eta=\frac{3}{4}\cos I_2 \left(\frac{a_{34}}{a_{12}}\right)^{3/2}\left(\frac{m_1+m_2}{m_3+m_4}\right)^{3/2}\ ,
\label{eqn:reson}
\end{equation}
where $\cos I_2$ is the inclination of the orbital plane of the second inner binary compared to the outer orbital plane. When $\eta\sim 1$, the timescale of the induced nodal precession and of the LK mechanism of the first binary are comparable and the resonance has its maximum effect.

\citet{liu2018} have recently discussed that dynamically induced BH mergers in quadruple systems may be an important channel of producing BH mergers observed by LIGO/VIRGO due to the richer dynamics of 2+2 systems.

\begin{table}
\caption{Models: name, maximum outer semi-major axis ($a_{\rm out}^{\max}$), slope of the BH mass function ($\beta$), semi-major axis distribution ($f(a)$), eccentricity distribution ($f(e)$), merger fraction ($f_{\rm mer}$).}
\centering
\begin{tabular}{lccccc}
\hline
Name & $a_{\rm out}^{\max}$ (AU) & $\beta$ & $f(a)$ & $f(e)$ & $f_{\rm mer}$\\
\hline\hline
A1 & $1000$ & $1$ & uniform     & uniform & $0.29$ \\
A2 & $1000$ & $2$ & uniform     & uniform & $0.29$ \\
A3 & $1000$ & $3$ & uniform     & uniform & $0.25$ \\
A4 & $1000$ & $4$ & uniform     & uniform & $0.24$ \\
B1 & $3000$ & $1$ & uniform     & uniform & $0.26$ \\
B2 & $1000$ & $1$ & log-uniform & uniform & $0.28$ \\
B3 & $1000$ & $1$ & uniform     & thermal & $0.35$ \\
\hline
\end{tabular}
\label{tab:models}
\end{table}

\section{N-Body Simulations: black hole mergers}
\label{sect:mergers}

\begin{figure} 
\centering
\includegraphics[scale=0.55]{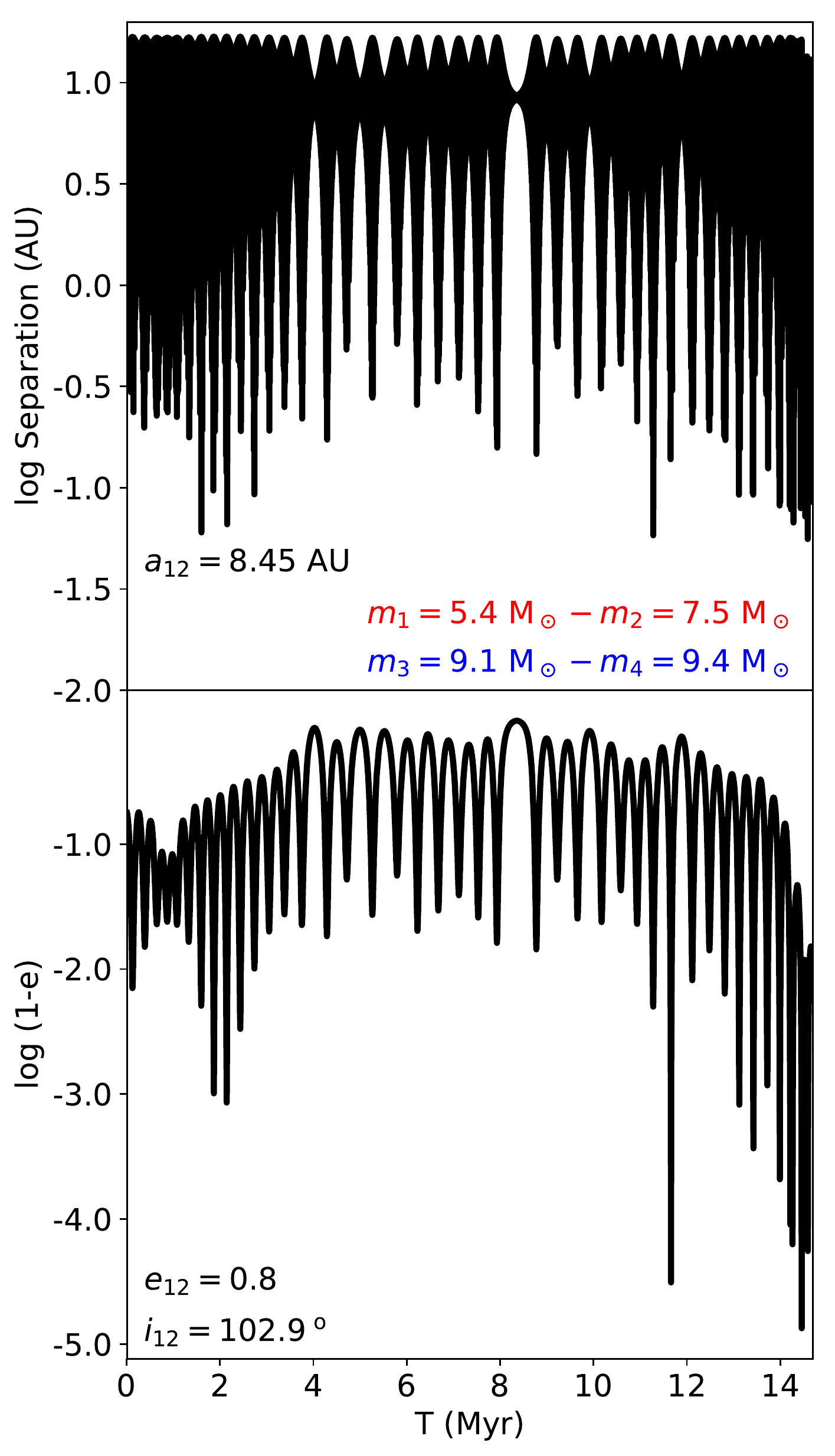}
\caption{Example of a four-body integration of one quadruple system with masses $m_1=5.4\msun$, $m_2=7.5\msun$, $m_3=9.1\msun$, $m_4=9.4\msun$. The first two BHs start with a semi-major axis $a_{12}=8.45$ AU, an eccentricity $e_{12}=0.8$ and an orbital inclination of $i_{12}=102.9^\circ$, and merge within $\sim 1.5\times 10^7$ yr.}
\label{fig:example}
\end{figure}

The BH quadruples in our simulations are initialized as follows. In total, we consider seven different models with different distributions of quadruple parameters (see Tab.~\ref{tab:models}) to explore the role of the relevant quadruple parameters.

In our models, we sample the masses of the BHs from
\begin{equation}
\frac{dN}{dm} \propto M^{-\beta}\ ,
\label{eqn:bhmassfunc}
\end{equation}
in the mass range $5\msun$-$100\msun$ \citep{hoan18}. To check how the results depend on the slope of the BH mass function, we run models with $\beta=1$, $2$, $3$, $4$ \citep{olea16}. The inner ($\aina$, $\ainb$) and outer ($\aout$) semi-major axes are drawn either from a uniform distribution ($f(a)\propto const$) or a log-uniform ($f(a)\propto 1/a$) distribution. The minimum inner semimajor axis is $1$ AU. We also consider two different maximum outer semimajor axis $\amax$
for the quadruples ($1000$ AU or $3000$ AU). The inner ($\eina$, $\einb$) and outer ($\eout$) eccentricities are sampled from a uniform ($f(e)\propto const$) or thermal ($f(e)\propto e$) distribution. The initial mutual inclination $i_0$ between the inner and outer orbit is sampled from an isotropic distribution (i.e. uniform in $\cos i$). The other relevant angles are drawn randomly.

After sampling the relevant parameters, we check that the initial configuration satisfies the stability criterion of hierarchical triples of \citet{mar01}, which is assumed to be valid for quadruple systems if the third companion is appropriately replaced by a binary system
\begin{equation}
\frac{R_{p}}{a_{\rm in}}\geq 2.8 \left[\left(1+\frac{m_{\rm out}}{m_{\rm in,1}+m_{\rm in,2}}\right)\frac{1+e_{out}}{\sqrt{1-e_{out}}} \right]^{2/5}\ .
\label{eqn:stabts}
\end{equation}
In the previous equation, $m_{\rm in,1}$ and $m_{\rm in,2}$ represent the masses of the inner binary, $m_{\rm out}$ the total mass of the binary companion and $R_p=\aout(1-\eout)$ is the pericentre distance of the outer orbit.

Given the above set of initial parameters, we integrate the system of differential equations of motion of the 4-bodies
\begin{equation}
{\ddot{\textbf{r}}}_i=-G\sum\limits_{j\ne i}\frac{m_j(\textbf{r}_i-\textbf{r}_j)}{\left|\textbf{r}_i-\textbf{r}_j\right|^3}\ ,
\end{equation}
with $i=1$,$2$,$3$,$4$, by means of the \textsc{ARCHAIN} code \citep{mik06,mik08}, a fully regularized code able to model the evolution of binaries of arbitrary mass ratios and eccentricities with high accuracy and that includes PN corrections up to order PN2.5. We performed $1000$ simulations for each model in Tab.~\ref{tab:models}. We fix the maximum integration time as \citep{sil17}
\begin{equation}
T=\min \left(10^3 \times \max\left(T_{\rm LK,{12}},T_{\rm LK,{34}}\right), 10\ \gyr \right)\,,
\end{equation}
where $T_{\rm LK,{12}}$ and $T_{\rm LK,{34}}$ are the LK timescales of the first ($m_1, m_2$) and second ($m_3, m_4$) inner binary, respectively.

\begin{figure} 
\centering
\includegraphics[scale=0.55]{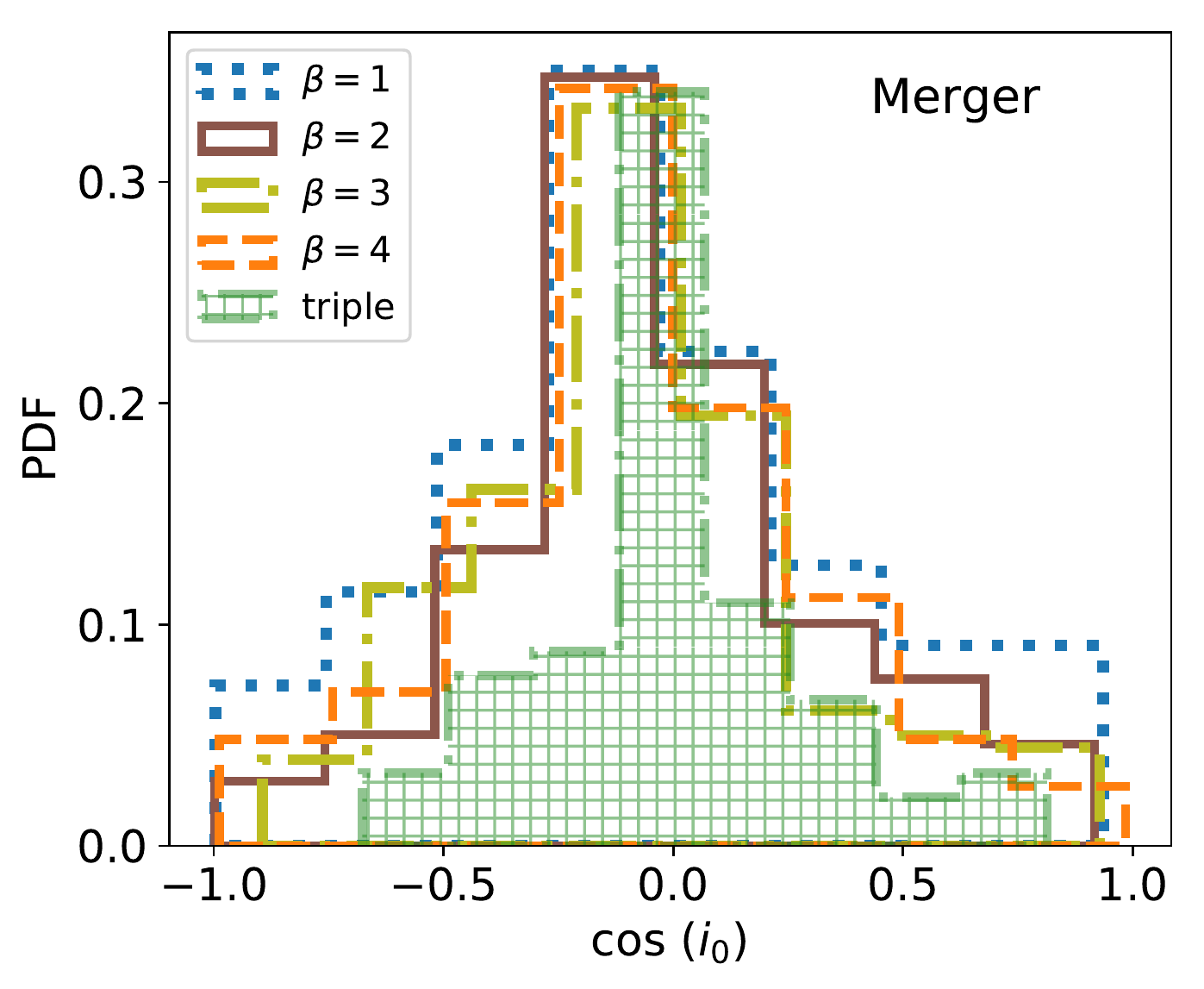}
\includegraphics[scale=0.55]{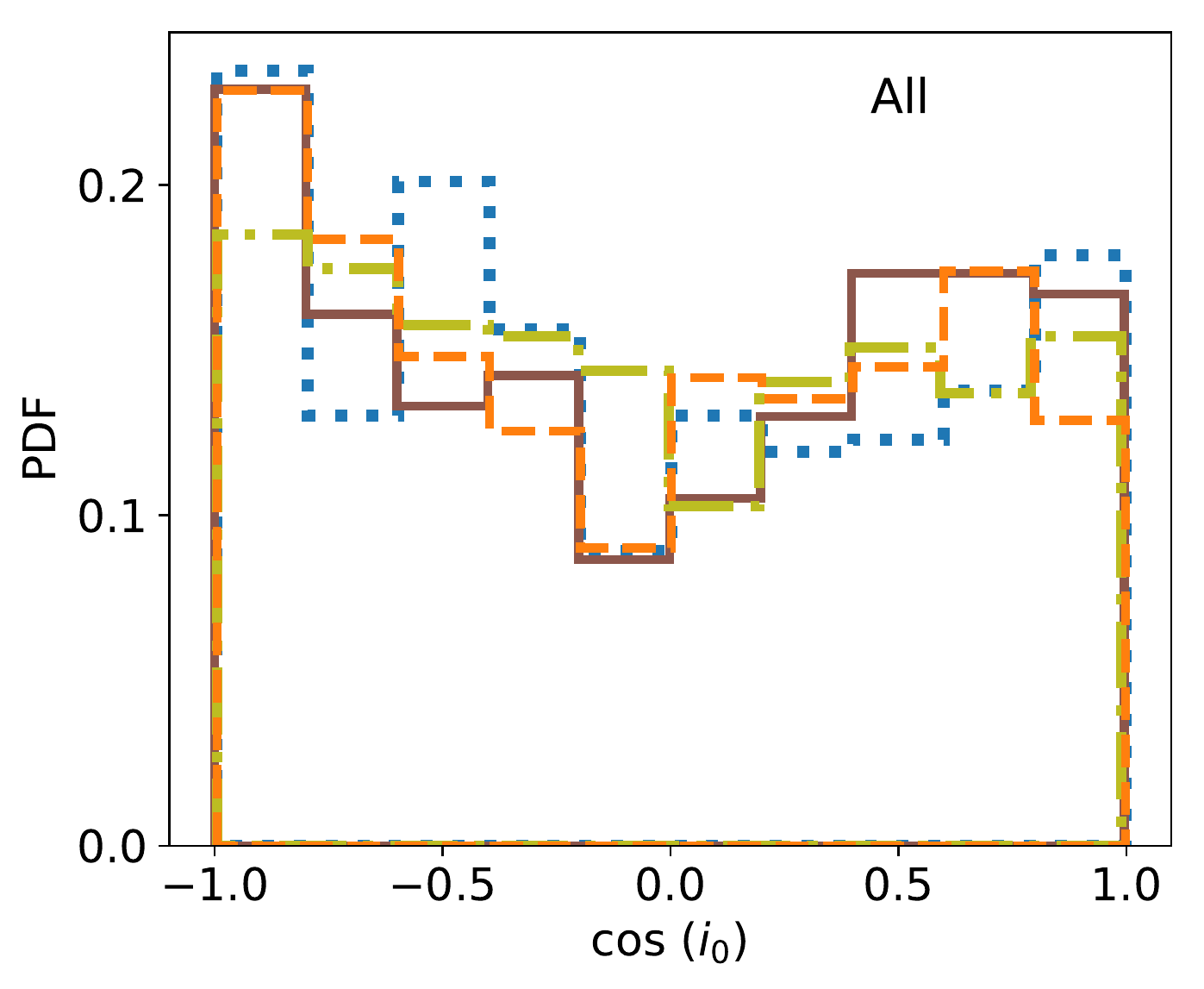}
\caption{Initial inclination distribution of BH binaries that merge (top) and that do not merge in the simulation (bottom). Compared to triple systems, excursions to very high eccentricity can take place over a much larger fraction of the parameter space and the distribution is broader.}
\label{fig:incl}
\end{figure}

\begin{figure} 
\centering
\includegraphics[scale=0.55]{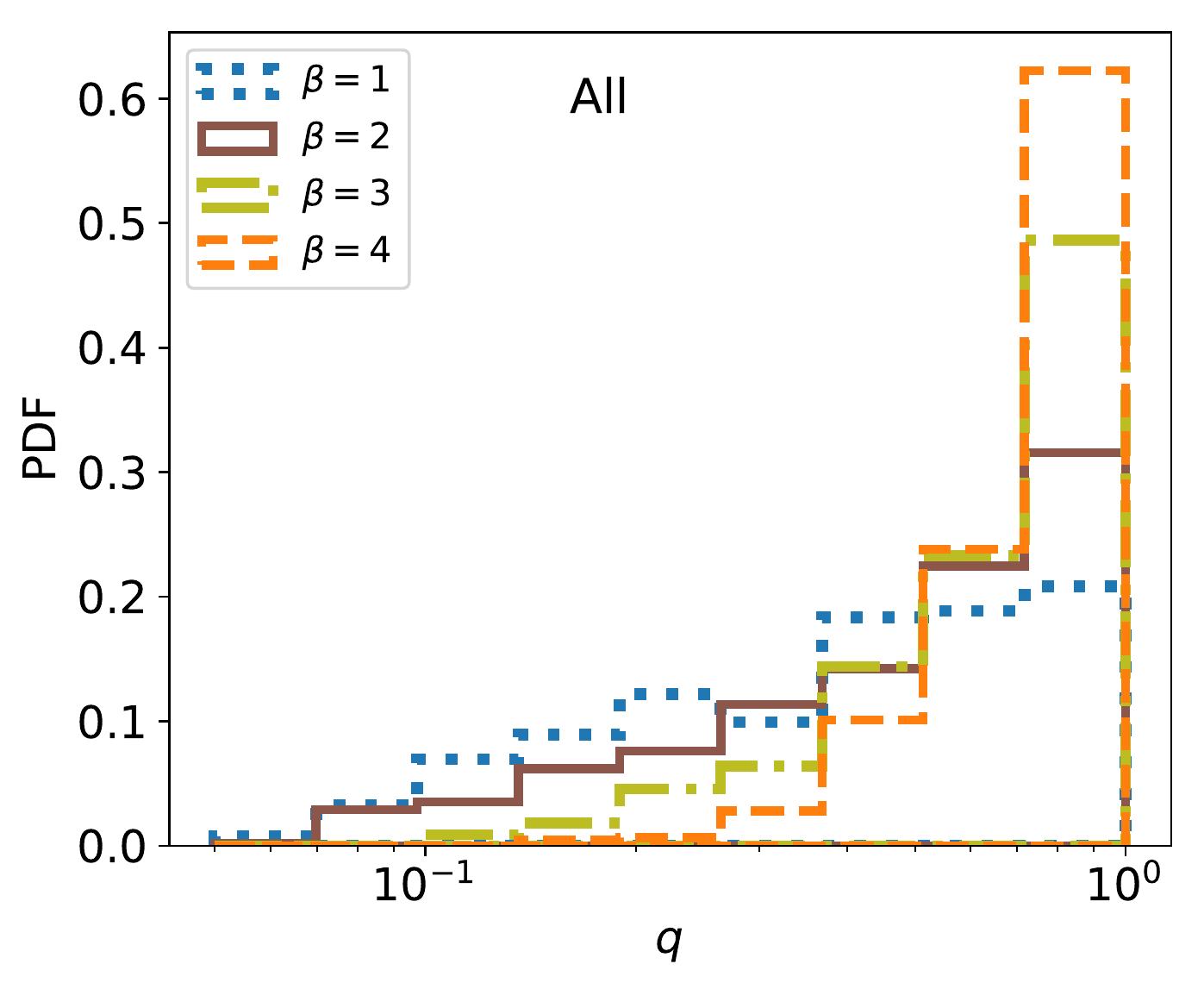}
\includegraphics[scale=0.55]{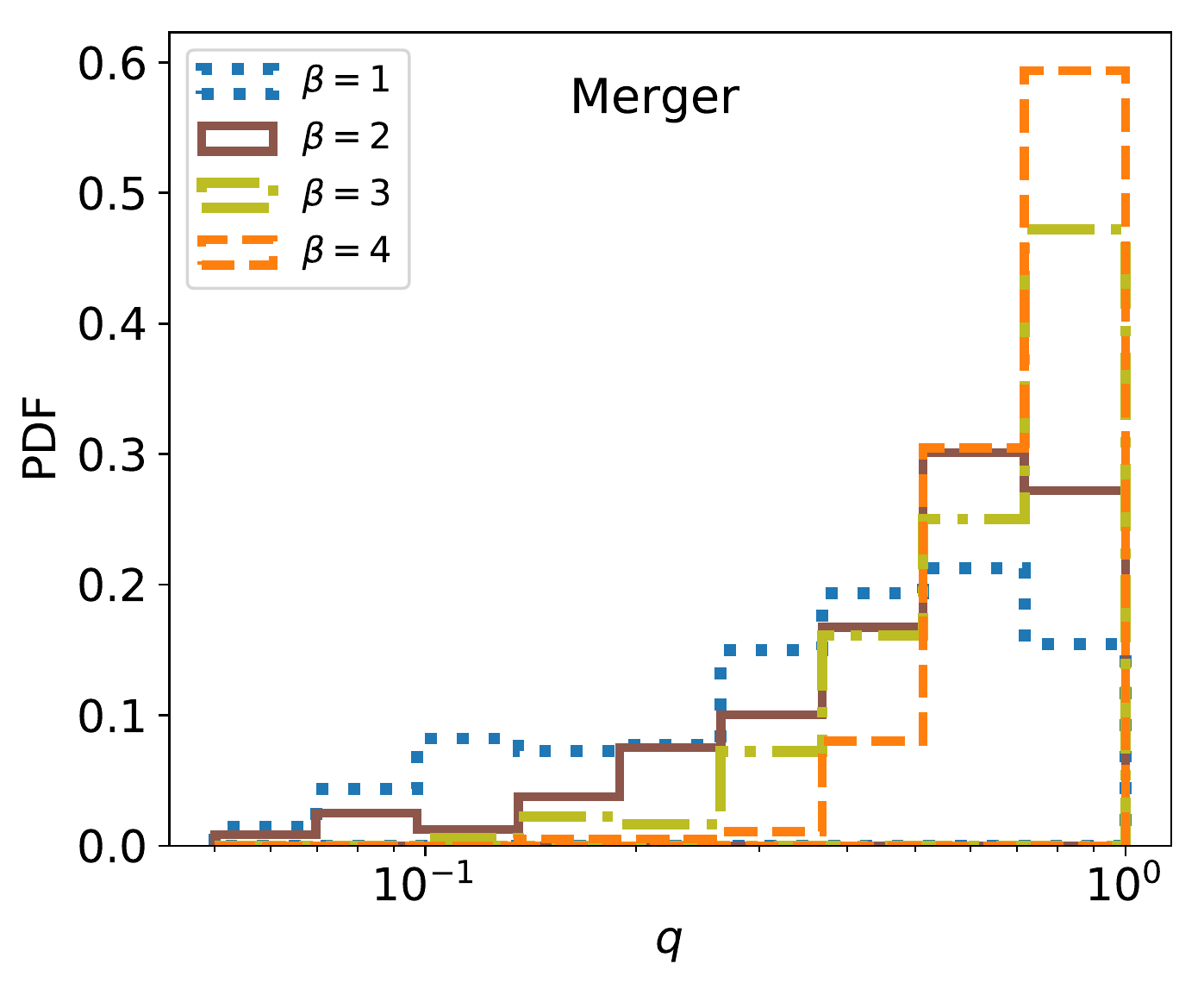}
\caption{Mass ratio distribution of BH binaries in quadruples in the simulation (top) and of BH binaries that merge (bottom) for different values of $\beta$. There is no significant enhancement of merger probability for any mass ratio in comparison to the prior mass ratio distribution.}
\label{fig:massr}
\end{figure}

\begin{figure*} 
\centering
\includegraphics[scale=0.55]{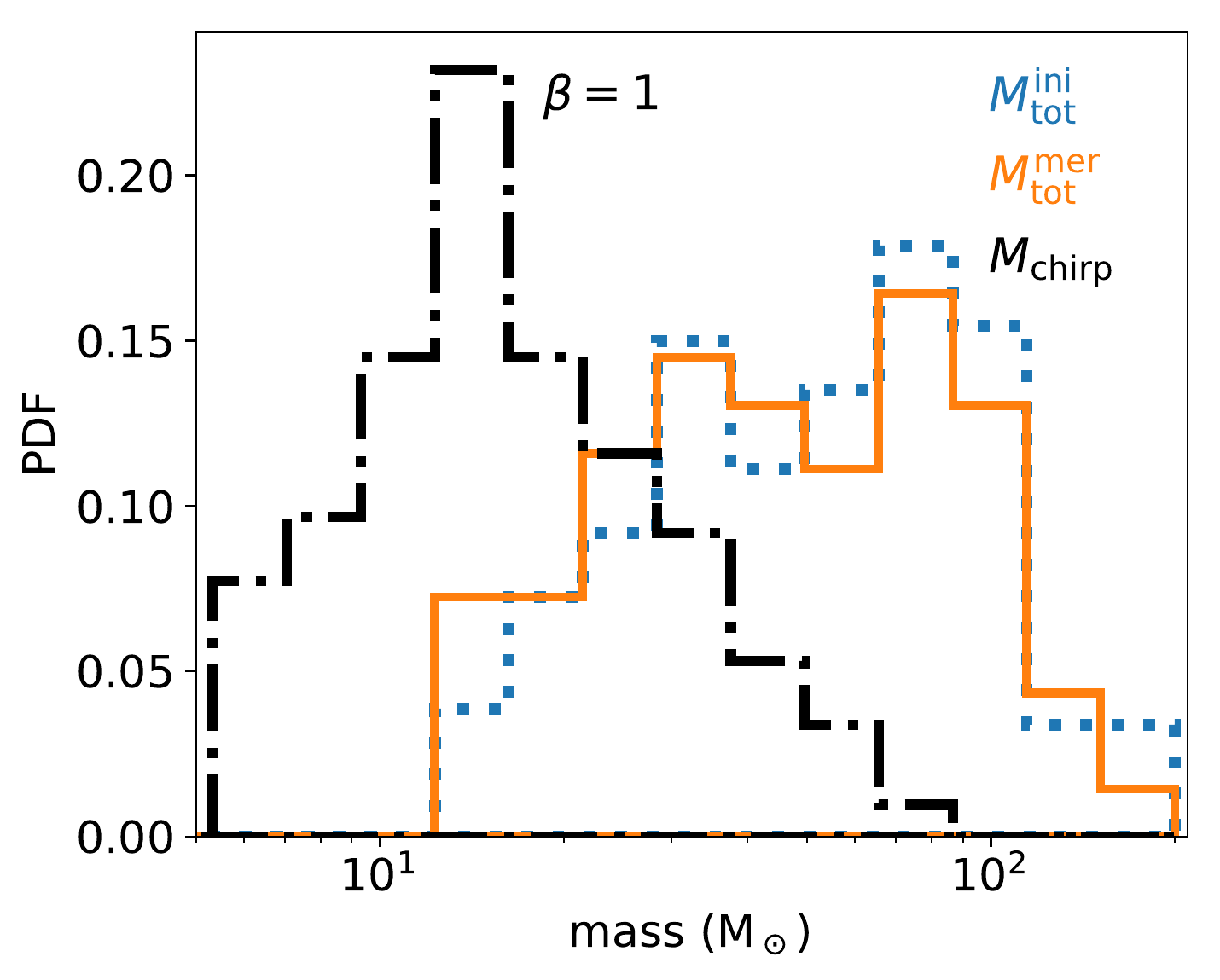}
\hspace{1.cm}
\includegraphics[scale=0.55]{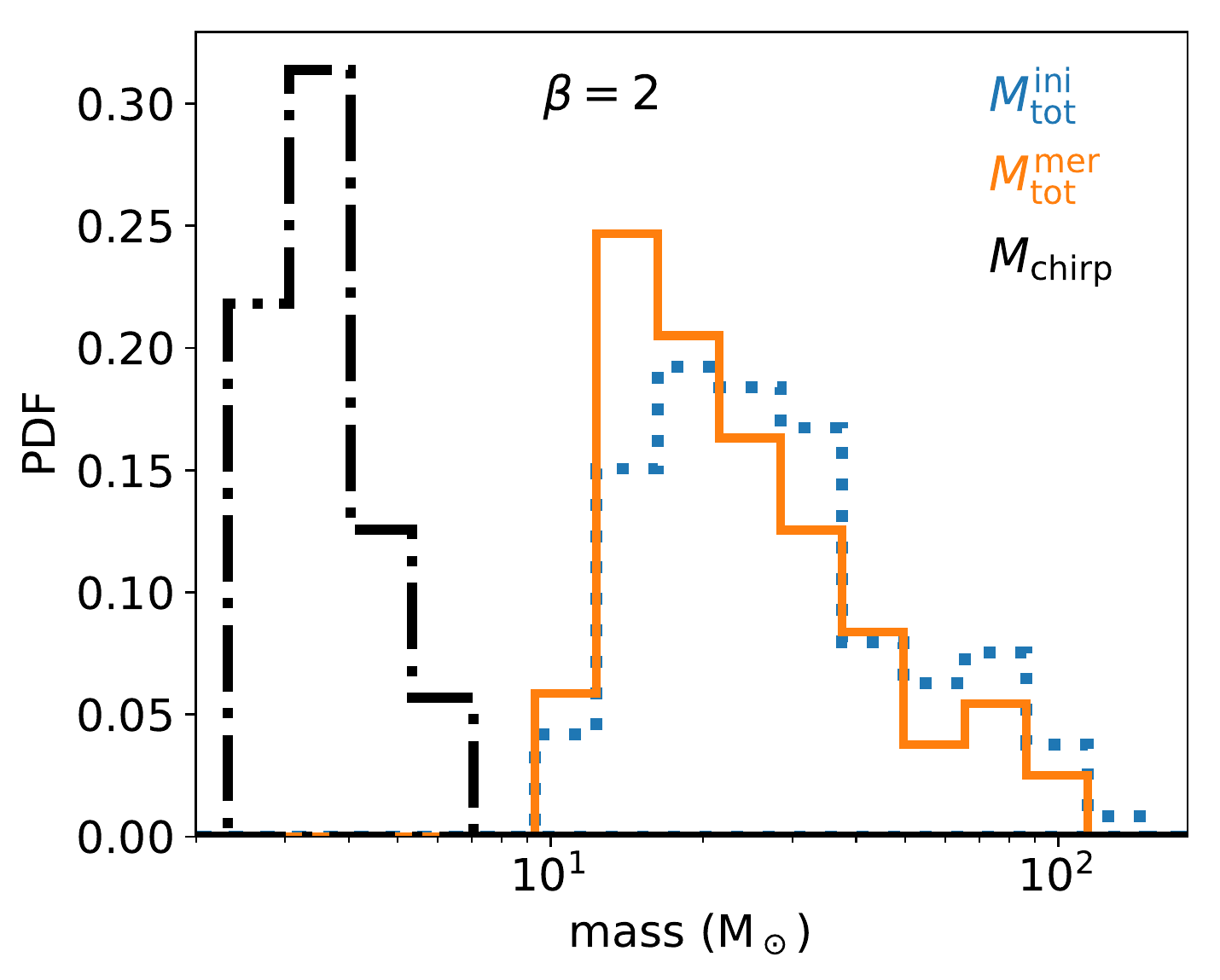}
\includegraphics[scale=0.55]{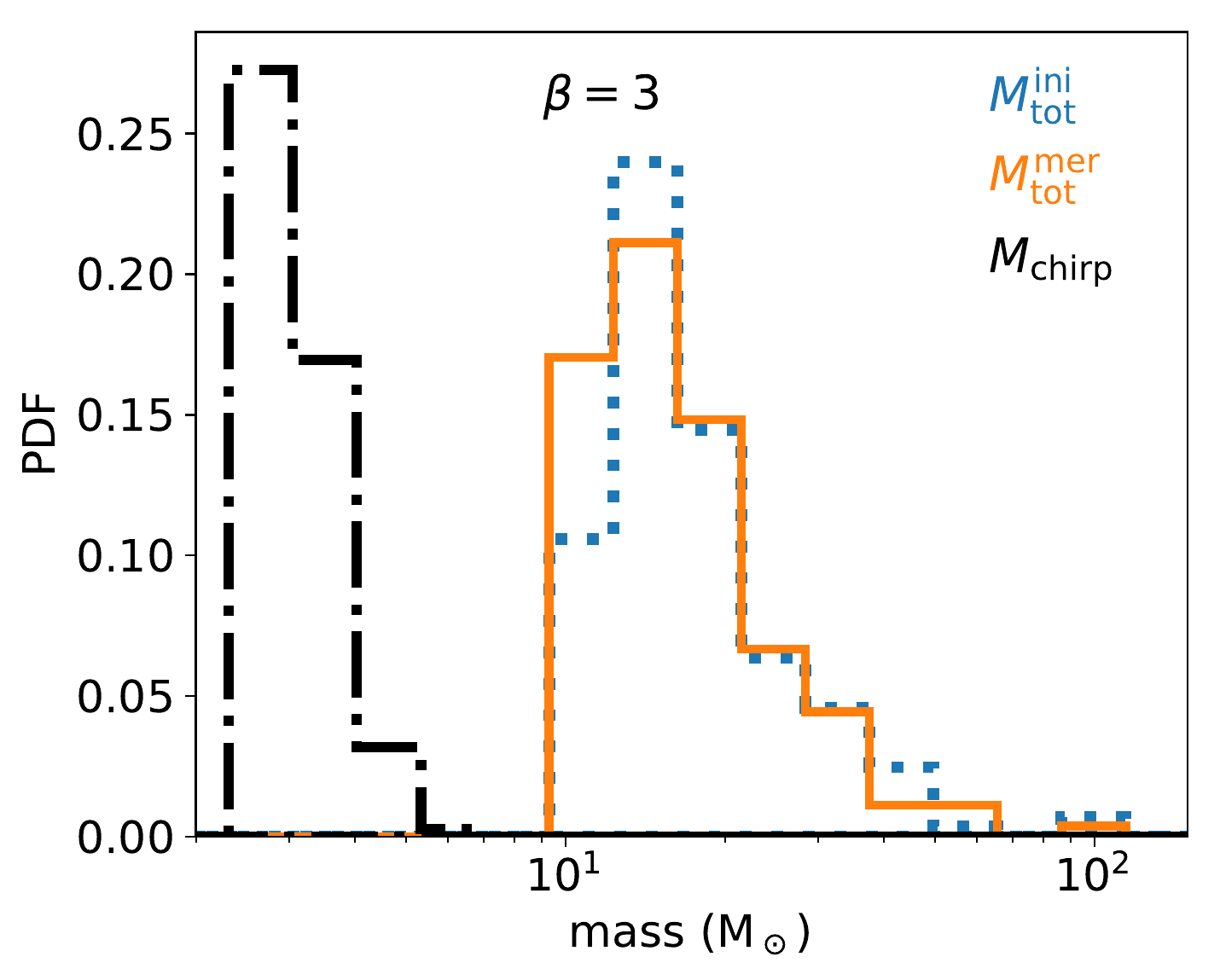}
\hspace{1.cm}
\includegraphics[scale=0.55]{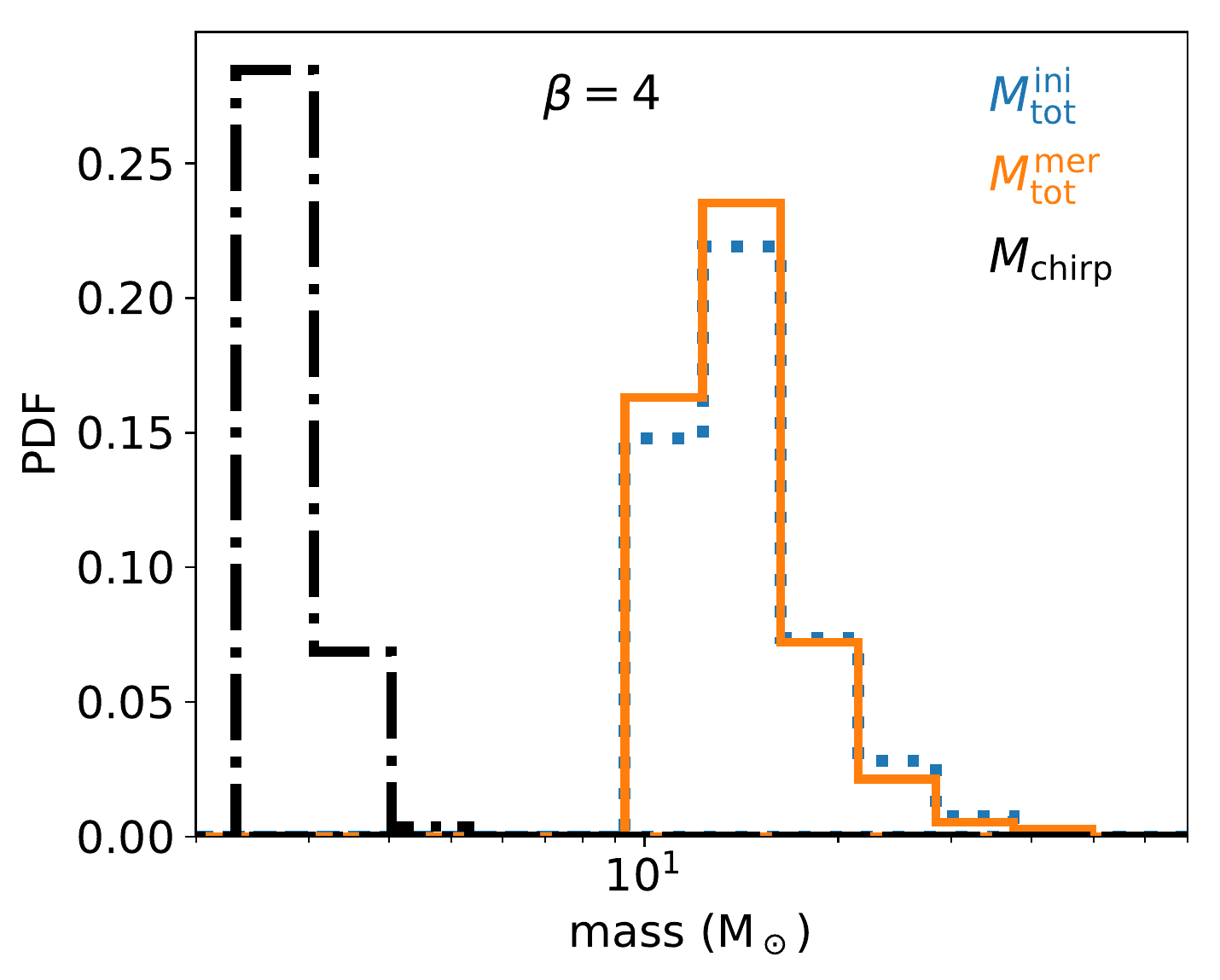}
\caption{Total mass distribution of BH binaries in quadruples, total mass of BH binaries that lead to a merger, and chirp mass for $\beta=1$ (top-left), $\beta=2$ (top-right), $\beta=3$ (bottom-left), $\beta=4$ (bottom-right).}
\label{fig:mass}
\end{figure*}

Figure~\ref{fig:example} shows the results of a four-body integration of one representative quadruple system with masses $m_1=5.4\msun$, $m_2=7.5\msun$, $m_3=9.1\msun$, $m_4=9.4\msun$. The first two BHs start with an orbital inclination of $i_{12}=102.9^\circ$ with respect to the quadruple orbital plane, semi-major axis $a_{12}=8.45$ AU, and eccentricity $e_{12}=0.8$. As a consequence of the efficient dissipation at high eccentricities pumped up by the LK effect (bottom panel), the BHs merge after $\sim 1.5\times 10^7$ yr. We note that binaries may enter the non-secular regime during the maximum of an LK oscillation and the secular equation of motions are a poor description of the quadruple dynamics. In this regime, the inner binary BHs can be driven to a merger before general relativistic effects suppress the LK oscillations.

We now describe the key physical properties of the BHs that merge in our models.

\subsection{Inclination distribution}

In Figure~\ref{fig:incl}, we show the probability distribution function (PDF) of the initial inclination between the inner binary orbit and the outer orbit of the BH binaries in quadruples that lead to a merger (top panel) and do not lead to a merger (bottom panel), for different slopes $\beta$ of the BH mass function (Models A). Figure~\ref{fig:incl} shows that the majority of the merging systems have initially high inclinations and that the distributions peak at $\sim 90^\circ$. However, binaries in quadruples can experience coherent eccentricity oscillations and excursions to very high eccentricity over a much larger fraction of the parameter space compared to triple systems \citep{pejcha2013} and possible resonances between the nodal precessions and the LK oscillations can arise \citep{hamerslai2017}. As a consequence, the inclination distribution of merging BHs in our four-body simulations shows broader tails compared to the BH mergers in triples. To show this, we have run an additional model where we replace in Model A1 one of the binaries with a single BH of equivalent total mass\footnote{In these runs, we adopt as maximum integration time $T=\min \left(10^3 \times T_{\rm LK,{12}}, 10\ \gyr \right)$, where $T_{\rm LK,{12}}$ is the LK timescale of the inner binary of the triple.}. Figure~\ref{fig:incl} shows that the distribution still peaks at $\sim 90^\circ$, but with less important tails compared to the quadruple case. In particular, we find that $\sim 50\%$ of the mergers in the triple case have initial inclination in the range $80^\circ$--$100^\circ$, while this fraction decreases to $\sim 30\%$ for a quadruple BH.

\begin{figure*} 
\centering
\includegraphics[scale=0.55]{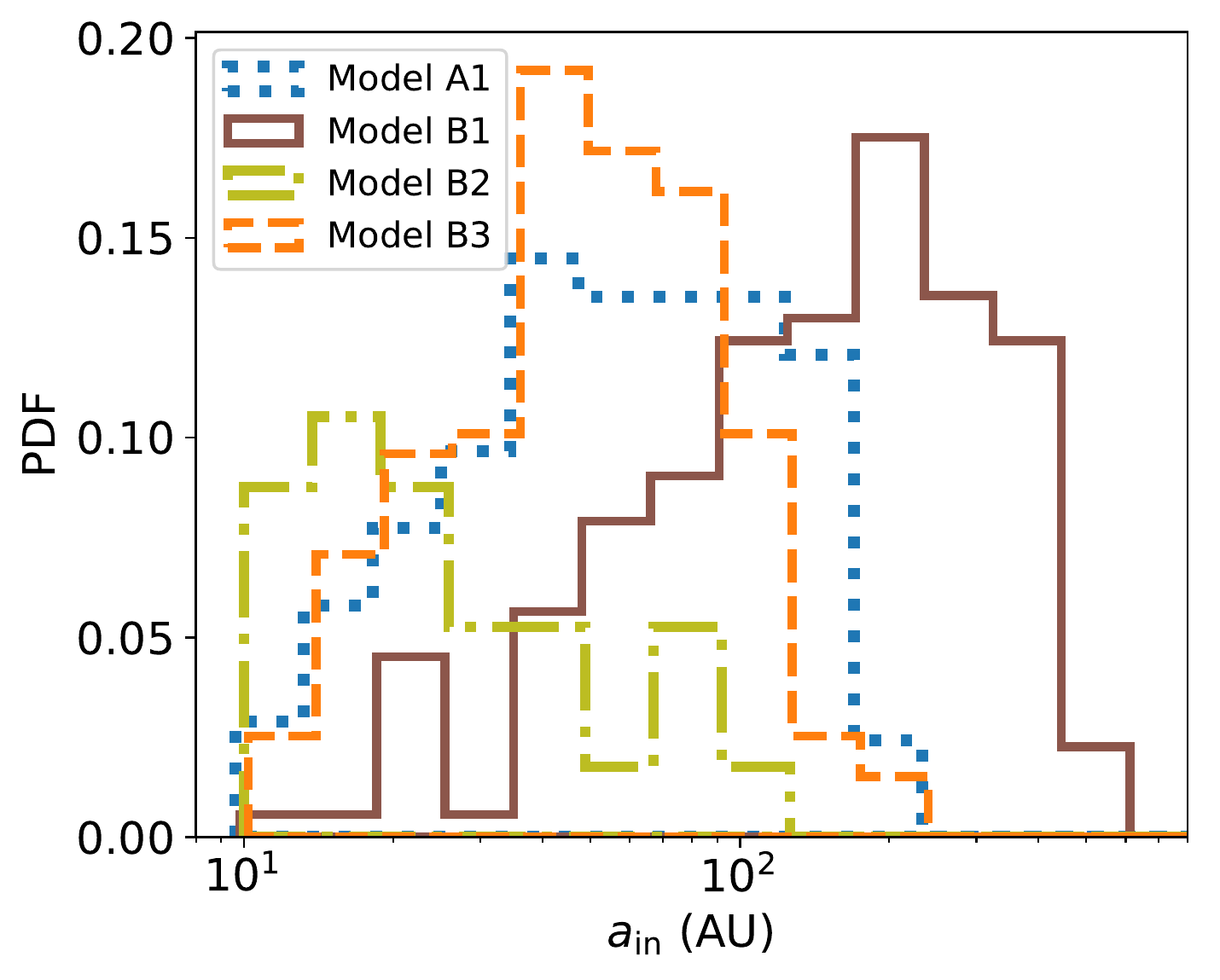}
\hspace{1.cm}
\includegraphics[scale=0.55]{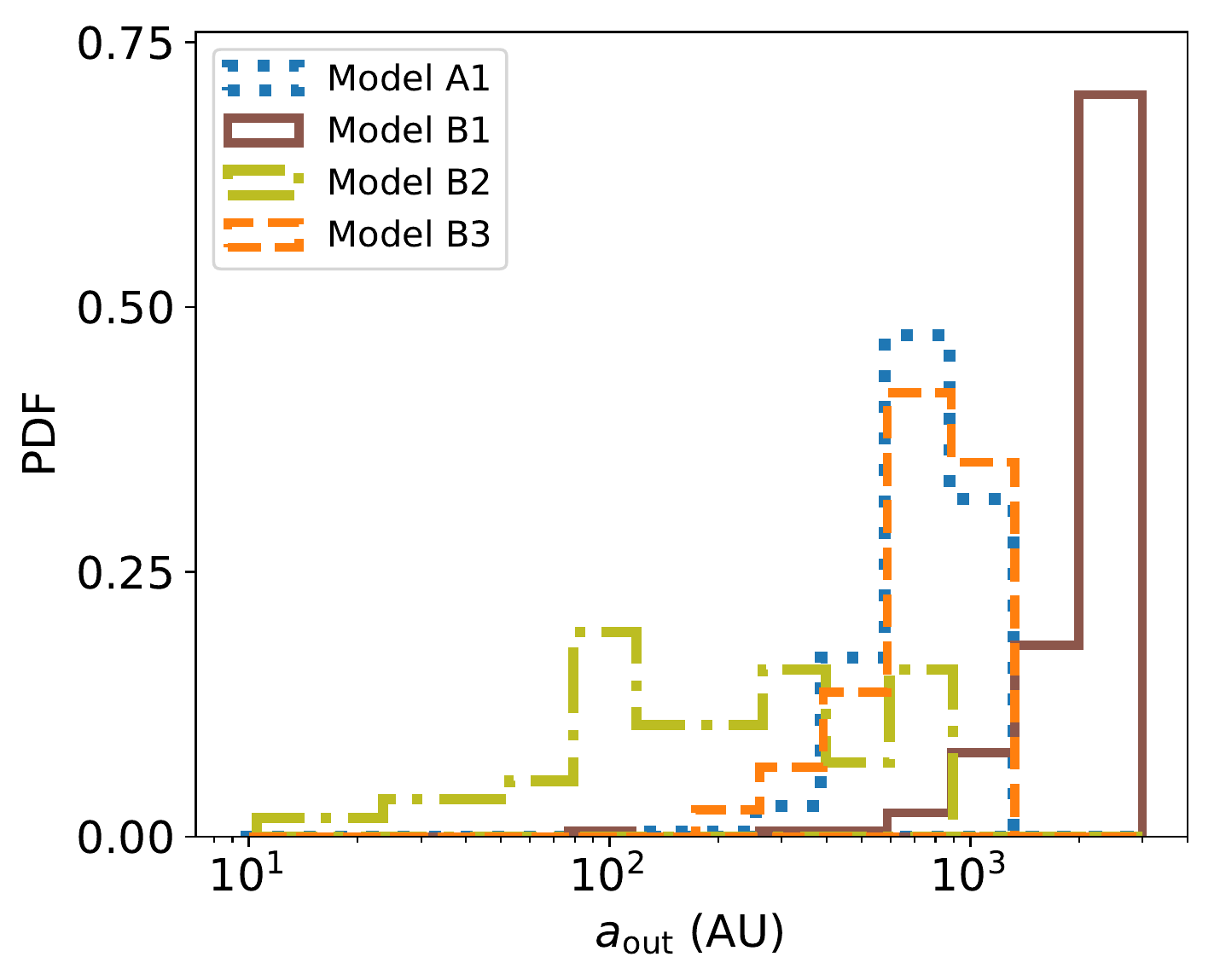}
\includegraphics[scale=0.55]{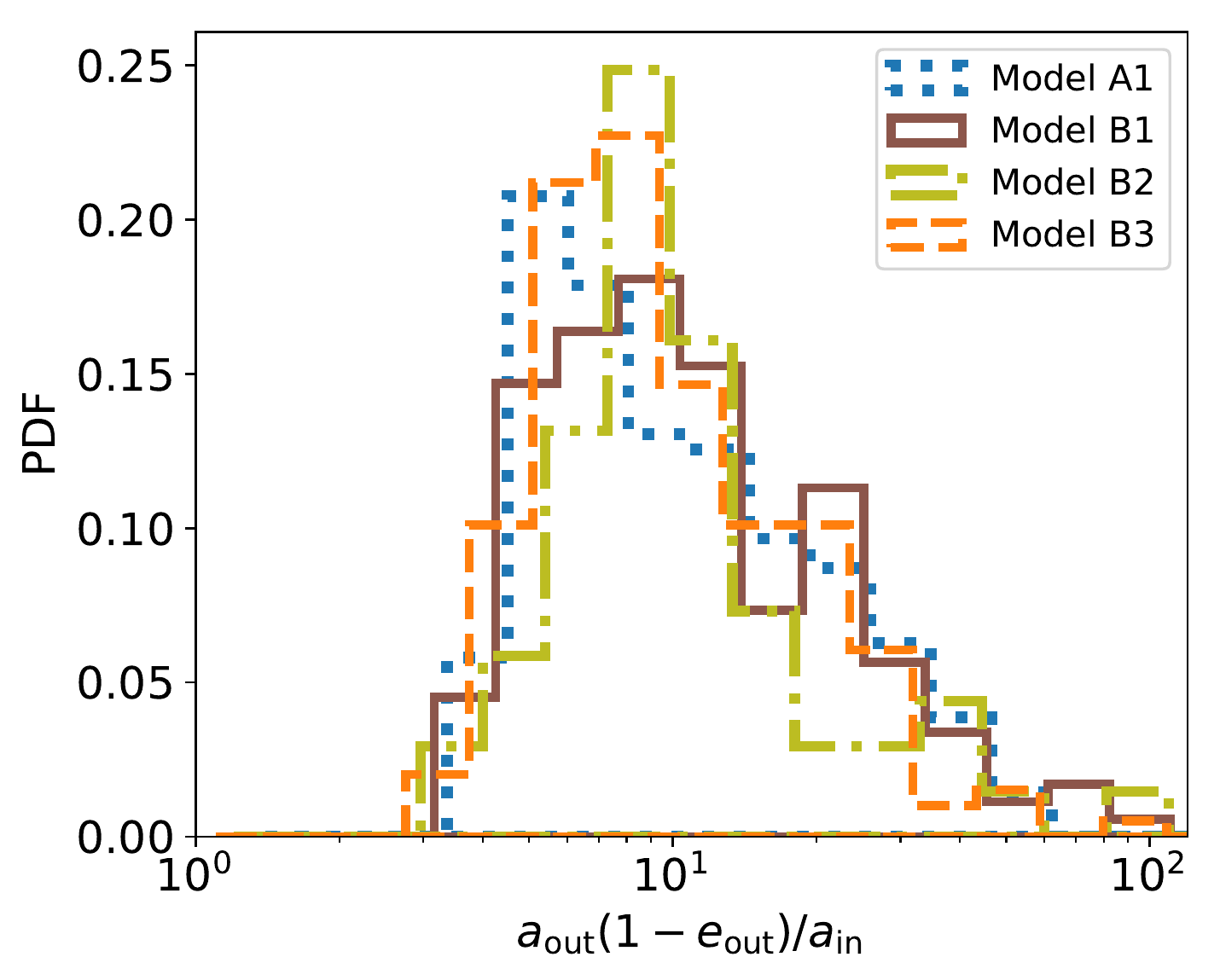}
\hspace{1.cm}
\includegraphics[scale=0.55]{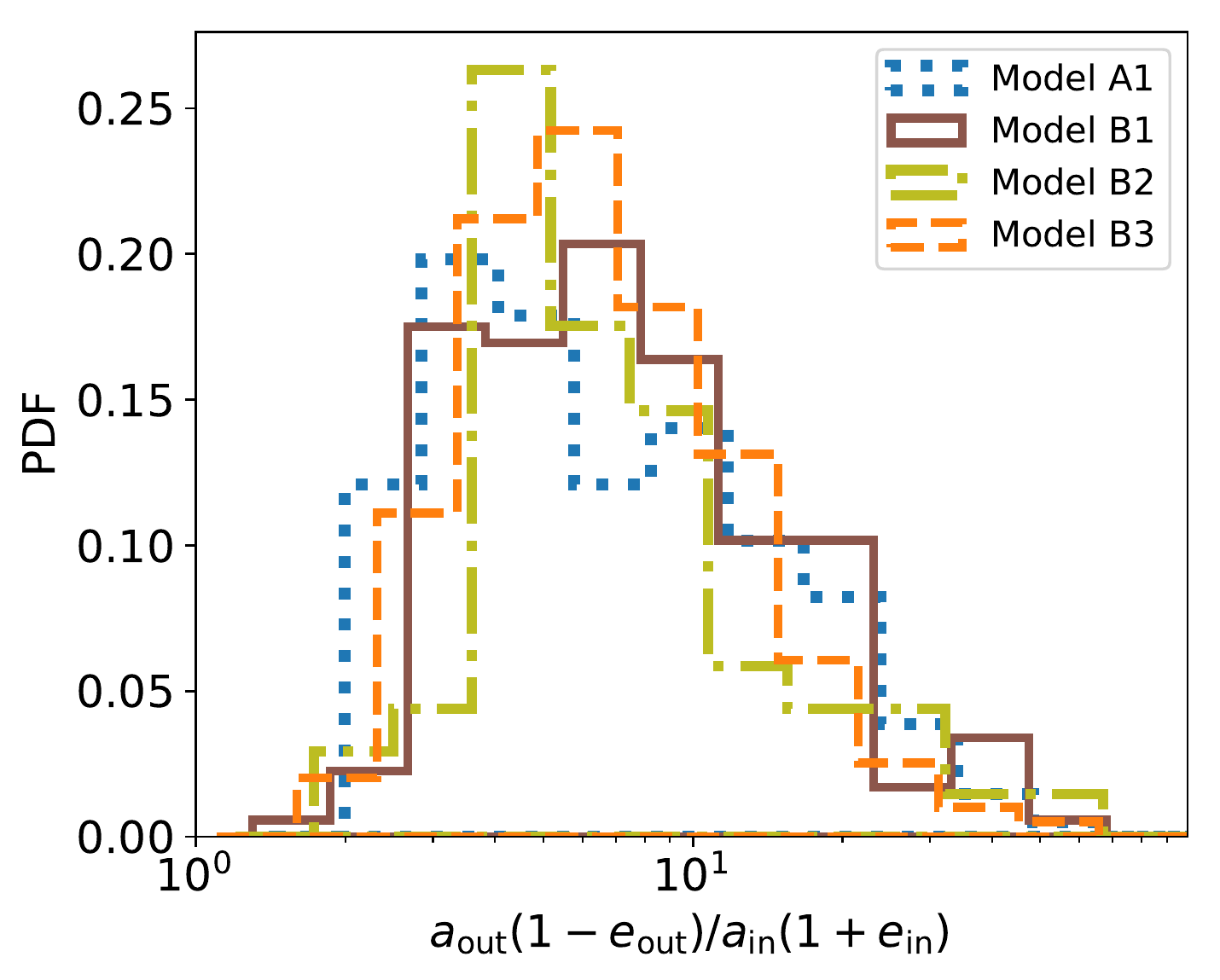}
\caption{Orbital distributions of of BH binaries in quadruples that lead to a merger for Models A1 and B (see Tab.~\ref{tab:models}): inner semi-major axis (top-left), inner eccentricity (top-right), outer semi-major axis (bottom-left), ratio of the outer pericentre to the inner semi-major axis (bottom-right).}
\label{fig:orbital}
\end{figure*}

\subsection{Masses}

To examine the expected total mass $m_{\mathrm{tot}}$ and the chirp mass
\begin{equation}
M_{\mathrm{chirp}}=m_{\mathrm{tot}}\frac{q^{3/5}}{(1+q)^{6/5}}
\end{equation}
of the merging binaries in quadruples ($q=m_2/m_1$, with $m_1>m_2$, is the mass ratio of merging BHs), we have run models with different values of the slope of the BH mass function.

Figure~\ref{fig:massr} reports the distribution of mass ratios of BH binaries in quadruples (top) and of BH binaries in quadruples (bottom) that lead to a merger for different $\beta$'s. We find that the distribution of mass ratios of merging BH binaries maps the overall distribution for all the values of $\beta$. Steeper mass functions predict mergers of BHs with similar masses, while shallower mass functions allow a broader distribution with mass ratio down to $q\sim 0.05$. The mass ratio distribution of merging systems follows the prior mass ratio distribution of randomly drawing two objects from the BH mass function.

Figure~\ref{fig:mass} reports the total mass distribution of BH binaries in quadruples, total mass of BH binaries that lead to a merger, and chirp mass for different $\beta$'s. Our merging binaries have total masses in the range $\sim 10$-$110\msun$, with the shape of the distribution that depends on the assumed slope of the BH mass function. Steeper mass functions (i.e. larger $\beta$'s) result in a smaller average $m_{\mathrm{tot}}$. We find that the total mass distribution of all binaries (including those which do not merge) is roughly constant for $\gtrsim 20\msun$ for $\beta=1$, while it is peaked at smaller total masses for larger $\beta$, i.e. at $\sim 20\msun$ for $\beta=2$ and $\sim 15\msun$ for $\beta=3$--$4$. We find that $\sim 90\%$ of the mergers have total mass $\lesssim 50\msun$, $\lesssim 30\msun$, and $\lesssim 20\msun$ for $\beta=2$, $3$, and $4$, respectively. For all $\beta$'s, we find that the distribution of the total mass of merging BH binaries tracks the underlying total mass distribution. Thus, there is no significant enhancement of the merger probability for any value of the total mass with respect to the prior distribution.

The distribution of chirp masses also depends on $\beta$, but the dependence is less sharp than the total mass. In the case $\beta=1$, we find that $M_{\mathrm{chirp}}$ is in the range $\sim 5$-$100\msun$ and its distribution presents a peak at $\sim 15\msun$. The peak of the distribution is at $\sim 3\msun$ and $M_{\mathrm{chirp}}$ is in the range $\sim 3$-$7\msun$ for $\beta=2$. Larger values of $\beta$ do not affect significantly neither the peak nor the shape of the chirp mass distribution.

\begin{figure*} 
\centering
\begin{minipage}{20.5cm}
\hspace{0.25cm}
\subfloat{\includegraphics[scale=0.55]{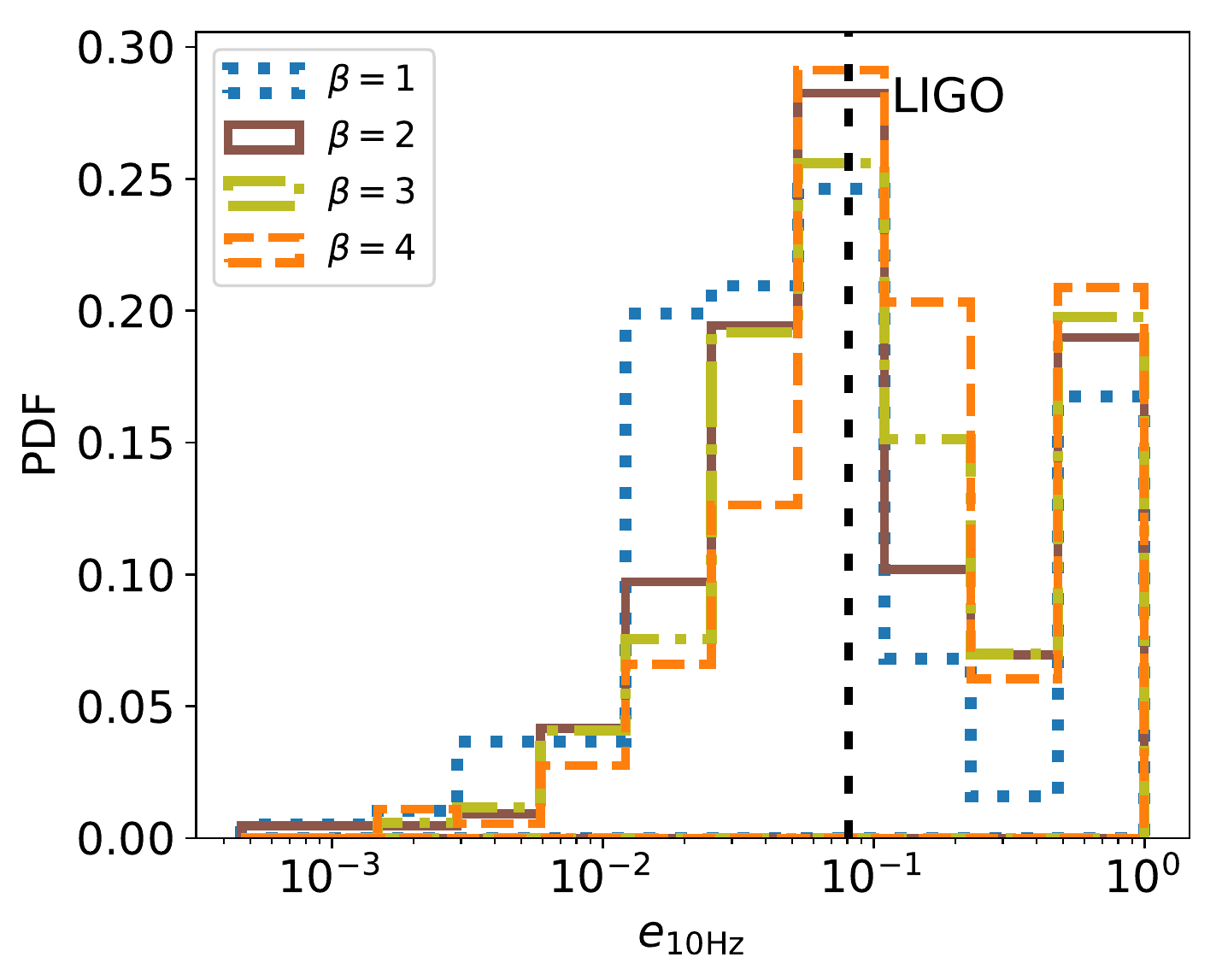}}
\hspace{1.cm}
\subfloat{\includegraphics[scale=0.55]{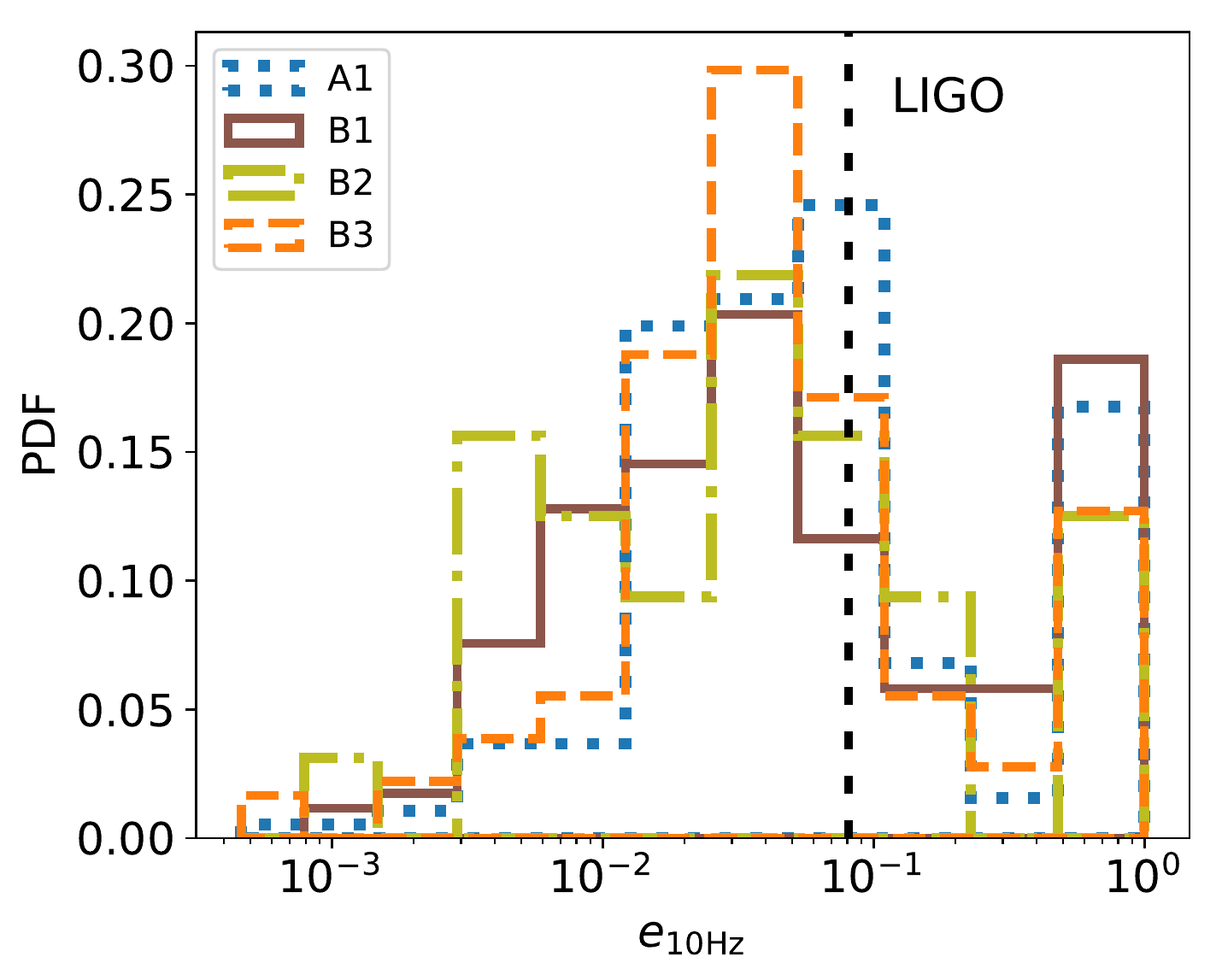}}
\end{minipage}
\caption{Distribution of eccentricities at the moment the BH binaries enter the LIGO frequency band ($10$ Hz) for mergers produced by quadruples. Left panel: Model A; right panel: Models A1 and B (see Tab.~\ref{tab:models}). The vertical line shows the minimum $e_{\rm 10Hz}=0.081$ where LIGO/VIRGO/KAGRA network may distinguish eccentric sources from circular sources \citep{gond2019}. A significant fraction of binaries formed in quadruples have a significant eccentricity in the LIGO band.}
\label{fig:ecc}
\end{figure*}

\subsection{Orbital semi-major axis and eccentricity}

Figure~\ref{fig:orbital} illustrates the distributions of orbital elements of the BH binaries in quadruples that lead to a merger for Model B.

In the top panel, we illustrate the distributions of inner (left) and outer (right) semi-major axes of merging binaries. The shape of the $\ain$ and $\aout$ distributions are mainly set by the initial distribution of semi-major axes and by the maximum extent of the quadruple $\amax$. In the case of Model A1 ($\amax=1000$ AU and $f(e)$ uniform), the distribution of inner semi-major axes is peaked at $\sim 50$ AU and extends up to $\sim 200$ AU. The distribution shifts to larger values with a peak at $200$ AU in Model B2, where the initial maximum semi-major axes was set to $3000$ AU, while it is peaked at $\sim 15$ AU in Model B2, where the initial semi-major axes are drawn from a log-uniform distribution. A similar trend is present in the distributions of outer semi-major axis, which are peaked at $\sim 900$ AU, $2500$ AU, $100$ AU for Model A1, Model B1 and Model B2, respectively.

The adopted distribution of initial eccentricities, i.e. either uniform (Model A1) or thermal (Model B3), does not affect the distributions of inner and outer semi-major axis of merging systems.

In the bottom panel of Figure~\ref{fig:orbital}, we plot the distribution of the ratio of the outer pericentre to the inner semi-major axis (left) and the distribution of the ratio of the outer pericentre to the inner apocentre (right) for merging systems. Both distributions are almost independent of the initial distribution of orbital semi-major axes and eccentricities. We find that the distributions peak at $a_{\rm out}(1-e_{\rm out})/a_{\rm in}\sim 10$ and $a_{\rm out}(1-e_{\rm out})/[a_{\rm in}(1+e_{\rm in})]\sim 5$ for merging binaries in quadruple systems universally for all of our models.

\subsection{Eccentricity in the LIGO band}

Figure~\ref{fig:ecc} shows the distribution of eccentricities at the moment the peak GW frequency of the BH binaries enter the LIGO frequency band ($f_{\rm GW}=10$ Hz) for mergers produced by quadruples, both for Model A (left) and Model B (right). Eccentric binaries emit a GW signal with a broad spectrum of frequencies. We compute the peak frequency of the GW spectrum as \citep{wen03}
\begin{equation} 
f_{\rm GW}=\frac{\sqrt{G(m_1+m_2)}}{\pi}\frac{(1+e_{\rm in})^{1.1954}}{[\ain(1-e_{\rm in}^2)]^{1.5}}\ .
\end{equation}

Figure~\ref{fig:ecc} shows that binaries formed in quadruples have larger eccentricities than those formed through many other channels, particularly mergers in isolated binary evolution and in SBHB ejected from star clusters \citep{frak18,gimap18,rod18}. However, mergers that follow from the GW capture scenario in clusters \citep{zevin18} and galactic nuclei \citep{olea09,gondan2018,rass2019}, from resonant binary-single scattering in clusters \citep{sam18}, from hierarchical triples \citep{ant17,fragrish2018,flp2019,fralo2019}, and from BH binaries orbiting intermediate-mass black holes in star clusters \citep{fragbr2019} also present a similar peak at high eccentricities. We note that the high eccentricities we find in the binaries that merge in our simulations may imply that a fraction of these binaries could emit their maximum power at higher frequencies, possibly in the range of LISA. However, binaries that enter the LIGO band with high eccentricities ($\gtrsim 10^{-3}$) could be harder to detect with instruments like LISA \citep{chen2017}. Nevertheless, \citet{randall2019} and \citet{hoang19} have proposed that LK oscillations of binary BHs in a triple configuration have the potential to be observed by LISA. In the quadruple case, both the inner BH binaries can enter the LISA frequency band and produce observable eccentricity variations if their respective LK cycles happen to have similar duration. 

\begin{figure} 
\centering
\includegraphics[scale=0.55]{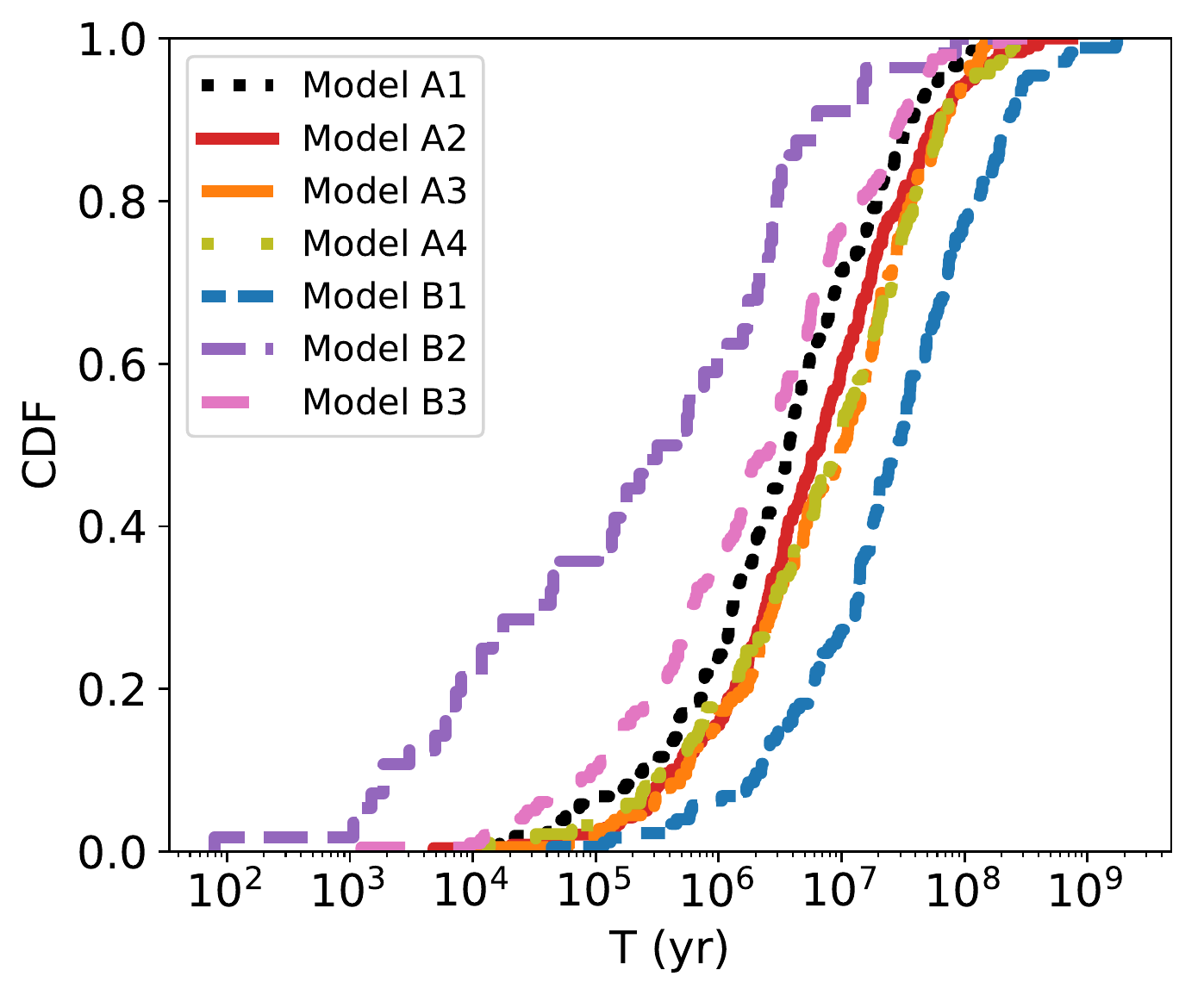}
\caption{Merger time distribution of BH binaries in quadruples that lead to merger for all models (see Tab.\ref{tab:models}).}
\label{fig:tmerge}
\end{figure}

\subsection{Merger times}

Figure~\ref{fig:tmerge} shows the merger time cumulative distribution functions (CDFs) of BH binaries in quadruples that lead to merger for Model A and Model B. The CDF function depends both on the maximum extent of the quadruple $\amax$ and on the initial distribution of semi-major axes, but it does not depend on the assumed slope of the BH mass function nor on the initial distribution of eccentricities. Larger $\amax$ imply a larger merger timescale. In Model A1 ($\amax=1000$ AU), we find that $\sim 50\%$ of the mergers happen within $\sim 10^7$ yr, while in Model B1 ($\amax=3000$ AU) within $\sim 10^8$ yr. In Model B2, where we sample the initial semi-major axes from a log-uniform distribution, $\sim 50\%$ of the BHs merge within $\sim 10^6$ yr.

\section{Discussion and conclusions}
\label{sect:conc}

In this paper, we have studied the dynamical evolution of a bound 2+2 type quadruple BH to estimate if this channel can contribute to the observed rate of BHs mergers. As found in \citet{liu2018}, quadruples have one additional degree of freedom and a richer dynamics, which can enhance the merger fraction.

We carried out a systematic statistical study of these systems by means of high-precision direct $N$-body simulations including Post-Newtonian (PN) terms up to 2.5PN order. We ran a series of simulations sampling a power-law mass spectrum for the BHs, different maximum extensions of the quadruple and different distributions of the inner and outer semi-major axes and eccentricities (Table~\ref{tab:models}). 
\begin{itemize}
 \item We found that the majority of the merging systems have high initial inclinations and that the distributions peak at $\sim 90^\circ$, as for triples, but with a broader distribution tail. 
 \item We have also shown that the mass distribution of merging BH binaries simply follows the assumed prior distribution, thus there is no significant enhancement of the merger probability as a function of total mass, chirp mass, or mass ratio. The slope of the BH mass function affects the total mass distribution and the chirp mass distribution: the steeper the mass function, the smaller the average $m_{\mathrm{tot}}$ and $M_{\mathrm{chirp}}$. Specifically, $\sim 90\%$ of the mergers have total mass $\lesssim 50\msun$, $\lesssim 30\msun$, $\lesssim 20\msun$ for $\beta=2$, $\beta=3$, $\beta=4$, respectively, assuming that the BH mass function extends to $100 \msun$.
 \item The shape of the inner and outer semi-major axes distributions of merging systems, as well as the typical merging timescale, are mainly set by the initial distribution of semi-major axes and by the maximum extent of the quadruple $\amax$. The initial eccentricity distribution of the inner binaries does not play a significant role. We find that the distributions peak at $a_{\rm out}(1-e_{\rm out})/a_{\rm in}\sim 10$ and $a_{\rm out}(1-e_{\rm out})/[a_{\rm in}(1+e_{\rm in})]\sim 5$ for merging binaries in quadruple systems universally for all of our models. 
 \item Finally, we have computed the eccentricity in the LIGO band and showed that BHs merging through this channel generate significant eccentricities. These are typically larger than BHs formed through mergers in isolated binary evolution and in SBHB ejected from star clusters \citep{frak18,gimap18,rod18}, but comparable to mergers that follow from the GW capture scenario in clusters \citep{sam18,zevin18} and galactic nuclei \citep{olea09,gondan2018,rass2019}, from hierarchical triples \citep{ant17}, and from BH binaries orbiting intermediate-mass black holes in star clusters \citep{fragbr2019}. Correlations between mass and eccentricity may distinguish some of these merger channels \citep{gondan2018}.
\end{itemize}

To directly compare to triples, we have also run an additional model where we replace in Model A1 one of the binaries with a single BH of equivalent total mass. We have found that the distribution of inclinations of BH binaries that lead to a merger is still peaked at $\sim 90^\circ$. However, $\sim 50\%$ of the mergers in the triple case have initial inclination in the range $80^\circ$--$100^\circ$, while this fraction decreases to $\sim 30\%$ for a quadruple BH. The distributions of masses, mass ratios and orbital elements of the merging BHs are not affected if one of the binaries in a hierarchical quadruple is replaced with a single BH in our simulations. Finally, we also investigated the role of PN terms by running Model A1 without PN corrections to the equations of motion. We have found that the fraction of BH binaries that merge decreases drastically. The reason is that there is no efficient energy dissipation due to GW emission at the pericenter when the eccentricity is excited as a consequence of the LK mechanism. The only available path for a merger is a head-on collision, when the two BHs reach a pericentre smaller than the sum of their Schwarzschild radii, typically of the order $\sim 10^{-7}$ AU. This means that, if the inner semi-major axis is $\sim 10$ AU, the LK mechanism should excite the eccentricity to $1-e_{\rm max}\sim 10^{-8}$, hence the initial inclination should be in the tiny interval $\sim 89.996$--$90.004$.

To accurately determine the global BH merger rate from quadruples, we would need to quantify the population of massive quadruple stars that lead to a quadruple BH of a given mass, which is currently poorly known. In our study, we have assumed a simple power-law distribution of BH masses in the range $5\msun$--$100\msun$.\footnote{Note that pulsational pair instability may limit the maximum mass to $\sim 50\msun$ \citep{bel2016}.} The multiplicity fraction of high-mass main-sequence stars, which may be the progenitors of NSs and BHs, can be as high as $\sim 80\%$ and quadruples are not rare, representing $\sim 20\%$--$25\%$ of the fraction of stellar objects in triples \citep{sana2017}. We find that the merger fraction in quadruples can be larger than that for triples, even up to $\sim 3$--$4$ times larger (see the last column of Tab.~\ref{tab:models}). Our results thus suggest that dynamically-driven BH and NS mergers in the quadruple scenario can be an important contribution to the events observed by LIGO/VIRGO.

\section*{Acknowledgements}

We thank Dong Lai for useful comments. GF thanks Seppo Mikkola for helpful discussions on the use of the code \textsc{archain}. GF is supported by the Foreign Postdoctoral Fellowship Program of the Israel Academy of Sciences and Humanities. GF also acknowledges support from an Arskin postdoctoral fellowship at the Hebrew University of Jerusalem. GF acknowledges hospitality from the E\"{o}tv\"{o}s Lor\'{a}nd University of Budapest. This project has received funding from the European Research Council (ERC) under the European Union's Horizon 2020 research and innovation programme under grant agreement No 638435 (GalNUC) and by the Hungarian National Research, Development, and Innovation Office grant NKFIH KH-125675 (to BK). Simulations were run on the \textit{Astric} cluster at the Hebrew University of Jerusalem.

\bibliographystyle{mn2e}
\bibliography{refs}

\end{document}